\documentclass[aps,prb,twocolumn,a4paper,showpacs,superscriptaddress]{revtex4-2}
\usepackage{graphicx,xcolor}% Include figure files
\usepackage{dcolumn}% Align table columns on decimal point
\usepackage{bm}% bold math
\usepackage{paralist}
\usepackage{hhline}
\usepackage{hyperref}% add hypertext capabilities

\begin{document}

\title{Fermi surface of the chiral topological semimetal CoSi}

\author{Nico Huber}
\email{nico.huber@tum.de}
\affiliation{Technical University of Munich, TUM School of Natural Sciences, Physics Department, D-85748 Garching, Germany}
\author{Sanu Mishra}%
\affiliation{Laboratoire National des Champs Magn\'{e}tiques Intenses (LNCMI-EMFL), CNRS, UGA, 38042 Grenoble, France}
\affiliation{Los Alamos National Laboratory, Los Alamos, New Mexico, 87545, USA}
\author{Ilya Sheikin}%
\affiliation{Laboratoire National des Champs Magn\'{e}tiques Intenses (LNCMI-EMFL), CNRS, UGA, 38042 Grenoble, France}
\author{Kirill Alpin}%
\affiliation{Max-Planck-Institute for Solid State Research, Heisenbergstrasse 1, D-70569 Stuttgart, Germany}
\author{Andreas P. Schnyder}%
\affiliation{Max-Planck-Institute for Solid State Research, Heisenbergstrasse 1, D-70569 Stuttgart, Germany}
\author{Georg Benka}%
\affiliation{Technical University of Munich, TUM School of Natural Sciences, Physics Department, D-85748 Garching, Germany}
\author{Andreas Bauer}%
\affiliation{Technical University of Munich, TUM School of Natural Sciences, Physics Department, D-85748 Garching, Germany}
\affiliation{Technical University of Munich, TUM Center for Quantum Engineering (ZQE), D-85748 Garching, Germany}
\author{Christian Pfleiderer}%
\affiliation{Technical University of Munich, TUM School of Natural Sciences, Physics Department, D-85748 Garching, Germany}
\affiliation{Technical University of Munich, TUM Center for Quantum Engineering (ZQE), D-85748 Garching, Germany}
\affiliation{Munich Center for Quantum Science and Technology (MCQST), D-80799 Munich, Germany}
\author{Marc A. Wilde}
\affiliation{Technical University of Munich, TUM School of Natural Sciences, Physics Department, D-85748 Garching, Germany}
\affiliation{Technical University of Munich, TUM Center for Quantum Engineering (ZQE), D-85748 Garching, Germany}

\date{May 6, 2024}% It is always \today, today,

%%%Abstract%%%
\begin{abstract}
We report a study of the Fermi surface of the chiral semimetal CoSi and its relationship to a network of multifold topological crossing points, Weyl points, and topological nodal planes in the electronic band structure.  Combining quantum oscillations in the Hall resistivity, magnetization, and torque magnetization with ab initio electronic structure calculations, we identify two groups of Fermi surface sheets, one centered at the R point and the other centered at the $\Gamma$ point. The presence of topological nodal planes at the Brillouin zone boundary enforces topological protectorates on the Fermi surface sheets centered at the R point. In addition, Weyl points exist close to the Fermi surface sheets centered at the R and the $\Gamma$ point. In contrast, topological crossing points at the R point and the $\Gamma$ point, which have been advertised to feature exceptionally large Chern numbers, are located at a larger distance to the Fermi level. Representing a unique example in which the multitude of topological band crossings has been shown to form a complex network, our observations in CoSi highlight the need for detailed numerical calculations of the Berry curvature at the Fermi level, regardless of the putative existence and the possible character of topological band crossings in the band structure. 
\end{abstract}

\maketitle

\section{Introduction}
Non-trivial topology in reciprocal space gives rise to a variety of interesting physical phenomena like Fermi arc surface states \cite{2011_Wan_PhysRevB, 2015_Xu_Science, 2015_Weng_PhysRevX, 2017_Chang_PhysRevLett}, the chiral anomaly \cite{2013_Son_PhysRevB, 2014_Burkov_PhysRevLett, 2015_Xiong_Science} and non-linear optical responses \cite{2016_Morimoto_PhysRevB, 2017_Chan_PhysRevB, 2017_deJuan_NatCommun, 2017_Wu_NaturePhys, 2020_Rees_SciAdv}. In recent years, chiral B20 compounds crystallizing in space group 198 have been intensively studied due to symmetry-protected topological degeneracies in their band structure including multifold crossing points \cite{2016_Bradlyn_Science, 2017_Tang_PhysRevLett, 2018_Pshenay-Severin_JPhysCondensMatter}, Weyl points and nodal planes \cite{2021_Wilde_Nature, 2022_Huber_PhysRevLett,2019_Yu_PhysRevB}. A prime example of this material class is the semimetal CoSi.

Angle-resolved photoemission spectroscopy (ARPES) \cite{2019_Rao_Nature, 2019_Sanchez_Nature, 2019_Takane_PhysRevLett} and quasiparticle interference \cite{2019_Yuan_SciAdv} experiments confirmed main features of the band structure predicted by density functional theory (DFT) \cite{2017_Tang_PhysRevLett,2018_Pshenay-Severin_JPhysCondensMatter} and found signatures of the topological crossing points by observing Fermi arc surface states. The optical properties were probed by investigating the optical conductivity \cite{2020_Xu_ProcNatlAcadSci} and measurements of the circular photo-galvanic effect (CPGE) \cite{2021_Ni_NatCommun}, further establishing the existence of exotic quasiparticles. However, the fine features of the Fermi surface pockets were not resolved in these studies.

A well-established method to experimentally determine the FS geometry with a high resolution is the analysis of quantum oscillations (QOs). Indeed, the electron pockets of CoSi around the R-point have been observed in detail in numerous QO studies \cite{2019_Wu_ChinPhysLett, 2019_Xu_PhysRevB, 2020_Wang_PhysRevB, 2022_Sasmal_JPhysCondensMatter} and the remarkably simple QO spectrum arising from this part of the FS has recently been explained comprehensively by the combined effect of nodal plane degeneracies and magnetic breakdown at near degeneracies \cite{2022_Guo_NatPhys,2022_Wilde_NatPhys, 2022_Huber_PhysRevLett}. These studies unequivocally confirm that the R-centered electron pockets occupy a Fermi volume that is considerably smaller than predicted by DFT. In contrast, little experimental evidence of the FS pockets around the $\Gamma$-point, which are predicted to enclose a multifold crossing \cite{2022_Huber_PhysRevLett}, has been reported. The only QOs related to the $\Gamma$-centered FS sheets detected experimentally are consistent with a single light and very small pocket \cite{2020_Wang_PhysRevB, 2022_Sasmal_JPhysCondensMatter}, which has not been assigned correctly to the band structure. The larger and heavier parts of the $\Gamma$-centered FS remain unobserved.

An experimental determination of the full FS is needed for the interpretation of response functions, in particular because charge neutrality in the semimetal CoSi together with the experimentally established volume of the R-centered pockets requires the Fermi volume of the predominantly hole-like $\Gamma$-centered pockets to differ from the DFT predictions as well. Such information on an experimentally refined FS is important for, both, an analysis based on topological charges alone and explicit numerical calculations of topological response functions, sampling the Berry curvature in non-trivial ways and naturally taking into account contributions from quantized topological charges and near degeneracies \cite{2004_Yao_PhysRevLett,2022_Guo_NatPhys,2022_Wilde_NatPhys} on an equal footing.
	
In this work, we report on the detection of QO frequencies in the Shubnikov-de~Haas (SdH) and de~Haas-van~Alphen (dHvA) oscillations of CoSi exhibiting heavy cyclotron masses. We systematically compare our findings with predictions from first-principles calculations allowing us to assign the experimentally observed oscillations to extremal orbits on the $\Gamma$-centered pockets and to determine their size, geometry and position with respect to the different topological crossings in their vicinity.

%%%%%%%%%%%%%%%%%%%%%%%%%%%%%%%%%%
\section{Electronic band structure}

\begin{figure*}
	\centering
	\includegraphics[width=\textwidth]{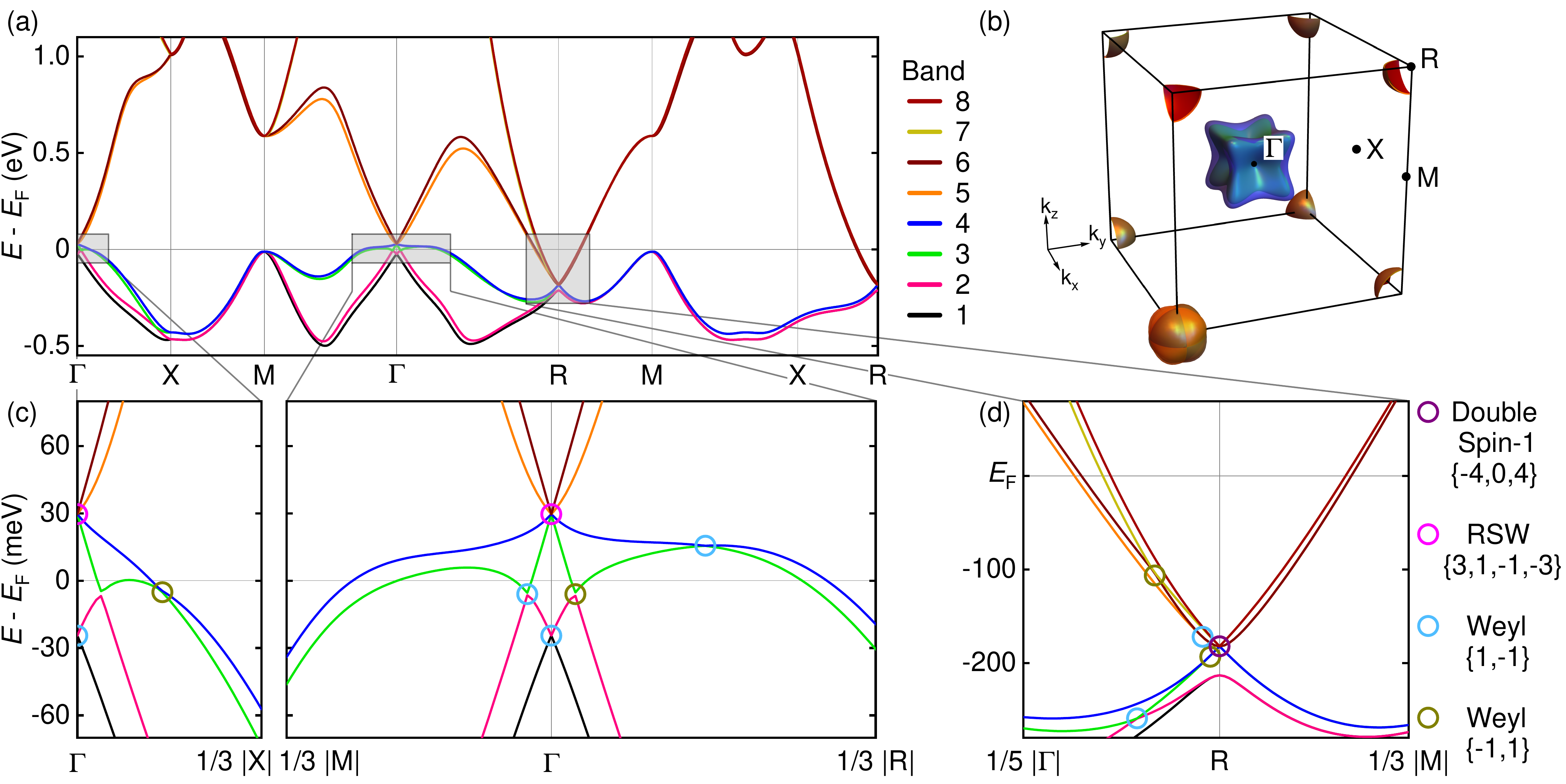}
	\caption{\label{fig:band structure-and-fs} Electronic structure of CoSi. (a)~Band structure calculated from first-principles including spin-orbit coupling. There are eight bands in the vicinity of the Fermi energy, which are pairwise degenerate at the Brillouin zone boundary. (b)~Fermi surface of CoSi comprising four almost spherical electron pockets centered at the R-point and multiple nested FS sheets around the $\Gamma$-point. The FS sheet arising from band 4 is depicted semi-transparent to allow for a view of the outer sheet arising from band 3. (c)~Detailed depiction of the bands along high-symmetry directions in the vicinity of the $\Gamma$-point with topological crossings marked by colored circles. Curly brackets denote the topological charge for each individual band (or band pair, where necessary) in ascending order. The 4-fold crossing at $\Gamma$ involving bands 3-6 represents a Rarita-Schwinger-Weyl fermion (RSW). Bands 1 and 2 exhibit a singular Weyl point at $\Gamma$. Further Weyl points occur on the $\Gamma$-X, $\Gamma$-M and $\Gamma$-R lines. (d)~Detailed views of the bands and crossing points around R. Note that the entire Brillouin zone boundary consists of nodal planes where all bands are pairwise degenerate, such that Chern numbers at R can only be defined for band pairs.
	}
\end{figure*}

The electronic band structure and the Fermi surface of CoSi were calculated with density functional theory (DFT) using \textsc{wien\oldstylenums{2}k}\cite{WIEN2k} in the generalized gradient approximation \cite{1996_Perdew_PhysRevLett}.
The effects of spin-orbit coupling were taken into account. The electronic structure was converged on a $23 \times 23 \times 23$ Monkhorst-Pack grid. Bands used for the determination of the Fermi surface were sampled on a $100\times 100\times 100$ k-mesh in the full Brillouin zone. For CoSi condensing in space group $P2_13$ (198) we used the experimental lattice constant $a=4.444$\,\AA. Both Co and Si occupy Wyckoff positions 4a with coordinates $(u,u,u),(-u+1/2,-u,u+1/2),(-u,u+1/2,-u+1/2),(u+1/2,-u+1/2,-u)$, where $u_{\rm{Co}}=0.143$ and $u_{\rm{Si}}=0.844$.
The band structure around the Fermi level comprising eight bands is shown in Fig.\,\ref{fig:band structure-and-fs}(a). Symmetry-enforced multifold crossing points at $\Gamma$ and R are located above and below the Fermi level, respectively.
 
Figure.\,\ref{fig:band structure-and-fs}(c) depicts the vicinity of the $\Gamma$-point. In addition to the multifold crossing points, there is a singular Weyl point at $\Gamma$ below the Fermi energy and multiple symmetry-enforced Weyl points along $\Gamma$-X, $\Gamma$-M, and $\Gamma$-R \cite{2022_Huber_PhysRevLett}. The distance of the symmetry-enforced crossings to the Fermi level as well as the question whether they are enclosed by FS pockets is beyond the accuracy of DFT. Fig.\,\ref{fig:band structure-and-fs}(d) shows the vicinity of the R-point, which displays a sixfold crossing point and Weyl points along high-symmetry directions well below the Fermi level as well as nodal planes on the Brillouin zone (BZ)
boundary. 

The curly brackets shown in Fig.\,\ref{fig:band structure-and-fs} denote the Chern numbers of each band involved in the crossing in the order of increasing band index. Chern numbers for single bands are well defined, e.g., $\{3,1,-1,-3\}$ for bands 3,4,5 and 6 at the $\Gamma$ point. In contrast, for the crossings associated with the R-point, no fully gapped integration contour may be defined due to the pairwise degeneracies at the nodal planes. In this case Chern numbers such as $\{-4,0,4\}$ may only be given  for the three pairs of bands $(3,4)$, $(5,6)$ and $(7,8)$, comprising the multifold crossing at R. Details concerning the calculation of Chern numbers are given in Ref. \onlinecite{2022_Huber_PhysRevLett}.
 
Shown in Fig.\,\ref{fig:band structure-and-fs}(b) is the calculated FS of CoSi, comprising four almost spherical electron pockets around the R-point and multiple nested sheets around the $\Gamma$ point. 

The shape and connectivity of the $\Gamma$-centered FS sheets are extremely sensitive to the precise location of the bands with respect to the Fermi level. Strong qualitative changes of the QO spectra make them an ideal tool to determine the FS of CoSi in the vicinity of the $\Gamma$-point. 

The main objective of the work reported in this paper concerns the experimental determination of the shape of the FS to lay the foundation for high-precision calculations of topological response functions. In CoSi such response functions are expected to depend sensitively on the Berry curvature at or around the FS.

%%%%%%%%%%%%%%%%%%%%%%%%%%%%%%%%%%%%%%
\section{Experimental Methods}

%%%%%%%%%%%%%%%%%%%%%%%%%%%%%%%%%%%%%%
\subsection{Sample growth}
High-quality single crystals of CoSi were grown using 99.995\% Co (MaTecK) and 99.9999\% Si (Alfa Aesar). Samples of similar quality were grown using either 99.9999\% Te (Alfa Aesar) as flux, or optical float zoning under ultra-high vacuum compatible conditions \cite{2011_Neubauer_RevSciInstrum}. For the flux growth, a mixture of Co, Si, and Te with an atomic ratio of 1:1:20 was loaded into a crucible and sealed in a quartz tube under Ar atmosphere. The mixture was heated in a rod furnace to 1150\,$^\circ$C, kept at this temperature for 20\,h, followed by slow cooling to room temperature at a rate of 3\,$^\circ$C/h. After removal of the Te flux using 50\% nitric acid, multiple millimeter-sized crystals with octahedral shape were obtained. For the optical float zoning polycrystalline rods were prepared from stoichiometric starting compositions of Co and Si using an inductively heated, ultra-high vacuum compatible rod-casting furnace \cite{2016_Bauer_RevSciInstrum}. The polycrystalline seed- and feed-rods were mounted in an optical mirror furnace and molten under high-purity Ar atmosphere resulting in large single-crystalline ingots.

Phase purity of the single crystals was confirmed by powder x-ray diffraction. The orientation of the single crystals was determined by Laue x-ray diffraction for both growth procedures. Thin platelets with edges parallel to major crystallographic axes were prepared using a wire saw. To further improve the crystalline quality, the platelets were annealed for 100\,h at 1100\,$^\circ$C. The annealing took place under Ar atmosphere for the flux grown sample and in ultra-high vacuum for the sample grown by optical float zoning. The residual resistivity ratios of the samples are 26 for the flux grown sample and 34 for the sample grown by optical float zoning.

%%%%%%%%%%%%%%%%%%%%%%%%%%%%%%%%%%%%%%
\subsection{Experimental setups}

Shubnikov--de Haas (SdH) oscillations were recorded in the Hall resistivity and transverse magnetoresistance under magnetic fields up to 18\,T. In this paper we focus on the Hall resistivity due to the larger absolute oscillation amplitude observed. The sample was contacted electrically by means of Al wire bonds. A small, low-frequency excitation current was applied and the transverse voltage pick-up recorded using a conventional lock-in technique. To improve the signal-to-noise ratio, the voltage pick-up was amplified by means of impedance matching transformers operated at liquid helium temperatures followed by suitable low-noise preamplifiers.

Using cantilever-based torque magnetometry with a capacitive readout, de Haas -- van Alphen oscillations were detected under applied magnetic fields up to 31.4\,T at LNCMI-Grenoble. These contributions to the torque originate in the non-linear field dependence of the dHvA effect and are a generic feature of materials exhibiting an anisotropic Fermi surface \cite{1965_Joseph_PhysRev,1966_Joseph_PhysRev,1970_Shoenberg_JLowTempPhys,1973_Griessen_Cryogenics,2008_Gasparov_PhysRevLett,2021_Siddiquee_JPhysCondensMatter,2019_Klotz_PhysRevB,1976_Ashcroft}. The magnetization under magnetic fields up to 7\,T was measured with a Quantum Design MPMS3 AC SQUID magnetometer.

For the magnetization measurements a large sample cut from the float-zoned single crystal was used to increase the signal-to-noise ratio. All other data presented in this paper were recorded on a flux grown sample. The SdH and cantilever-based dHvA measurements were conducted down to milli-Kelvin temperatures using dilution refrigerators, permitting in-situ sample rotation. The magnetization was measured down to 1.8\,K for fixed sample orientation.

%%%%%%%%%%%%%%%%%%%%%%%%%%%%%%%%%%%%%%
\subsection{Data analysis}

The data analysis used to infer the quantum oscillation spectra of the Hall resistivity and magnetization  involved the following steps (a pedagogical account may be found in the supplementary information of Ref.\,\cite{2022_Huber_PhysRevLett}). First, the nonoscillatory signal component was determined by fitting and subtracting polynomials between $2^{nd}$ and up to $5^{th}$ order. The order of polynomials was chosen to be as low as possible to avoid artifacts in the Fourier spectra, but as high as necessary to remove the background efficiently. The oscillatory signal component was interpolated on a set of evenly spaced $1/B$ values and multiplied by a windowing function to reduce spectral leakage. Zero-padding was employed to increase the FFT sampling rate. The peaks in the FFT spectra were fitted with Gaussians to determine oscillation frequencies and amplitudes.

The Hall resistivity and magnetization were recorded for both increasing and decreasing magnetic field at each temperature and field orientation. The variance in the QO frequencies and amplitudes observed in both sweep directions was used to determine the error bars shown. For the high-field dHvA data only the up- or downsweep of the magnetic field was recorded at a given temperature and angle. As an estimate of the uncertainties of the peak amplitudes detected, the noise level of the FFT spectra at a frequency close to the peaks was used.

We considered the following criteria in the identification of peaks in our FFT spectra: (i)~Where available, the data obtained under increasing and decreasing magnetic field were directly compared to each other. Only peaks that were detected for both sweep directions were considered further. (ii)~Following systematic subtraction of nonoscillatory signal contributions from the raw data approximated by polynomials of different degree, the calculated FFTs were compared. Only peaks that were consistently detected under different analysis parameters were considered as possible QO frequencies. (iii) To rule out artifacts due to spectral leakage, different windowing functions were applied before performing the FFT analysis. Only features that were consistently present for different windowing functions were accepted as oscillation frequencies. (iv)~Data were analyzed for different field ranges while keeping the window size in $1/B$ the same. Only peaks for which the FFT amplitude increased with increasing fields were considered as possible QO frequencies. (v)~We compared the data for different field directions systematically. Only oscillatory components present under several adjacent field directions were considered.

The temperature dependence of the quantum oscillation amplitude was analyzed by means of the temperature reduction factor $R_T$ in the Lifshitz-Kosevich formalism \cite{1956_Lifshitz_SovPhysJETP}
\begin{equation}\label{eq:RT}
	R_T=\frac{X}{\sinh(X)}
\end{equation}
with
\begin{equation}
	X=\frac{2\pi^2 p m^* k_B T}{\hbar e B},
\end{equation}
where $p$ is the harmonic number, $m^*$ the cyclotron mass, $k_B$ the Boltzmann constant, $T$ the temperature, $\hbar$ the reduced Planck constant, $e$ the electron charge and $B$ the magnetic flux density. As the Fourier analysis was carried out over a finite range in $1/B$, the mean value of the analysis window in $1/B$ was used in the evaluation. We estimate the overall uncertainty of the extracted cyclotron masses to be $\pm5\%$.

%%%%%%%%%%%%%%%%%%%%%%%%%%%%%%%%%%%%%%
\subsection{Predicted QO amplitudes}\label{App:QOAmps}
The relative strengths of the calculated frequency branches as depicted in terms of the linewidths shown in Fig.~\ref{fig:comparison-exp-theory} were calculated as follows. The cyclotron mass $m^*$ and the curvature of the cross-sectional area with respect to the field direction were inferred from the DFT data using the SKEAF tool \cite{2012_Rourke_ComputerPhysicsCommunications}. The expected relative QO amplitudes $A$ were calculated as
\begin{equation}\label{eq:AmplitudeFactor}
	A=\frac{f}{m^* \sqrt{c}} R_T R_D,
\end{equation}
where $f$ is the QO frequency, $c$ the curvature, and $R_T$ the temperature damping factor as given in Eq.~\ref{eq:RT}. $R_D$ is the Dingle damping factor given by
\begin{equation}
	R_D=\exp\left(-\frac{\pi p}{\tau \omega_{c}}\right)=\exp\left(-\frac{\pi p \sqrt{\frac{2 \hbar f}{e}}}{l_\mathrm{mfp} B}\right),
\end{equation}
where $\tau$ is the relaxation time, $\omega_{c}$ the cyclotron frequency, $l_\mathrm{mfp}$ the mean free path under the assumption of a circular cross section. Values of $B$=18\,T, $T$=0.1\,K and $l_\mathrm{mfp}$=30\,nm were used for the evaluation \onlinecite{CommentMFP}. For clarity, the width of the branch with a QO frequency of $\approx$20\,T, which displays a very low cyclotron mass as compared to the other frequency branches, is reduced by a factor of 10.

For the frequency branches between 600 and 2000\,T arising from the outer FS sheets of bands 3 and 4, which we only detect in our high-field experiments using cantilever magnetometry, the torque factor
\begin{equation}
	R_\mathrm{Torque}=\frac{1}{f}\frac{df}{d\theta}
\end{equation}
has been multiplied to the right-hand side of Eq.~\ref{eq:AmplitudeFactor}.

%%%%%%%%%%%%%%%%%%%%%%%%%%%%%%%%%%%%%%
\section{Experimental Results}

\begin{figure}
	\includegraphics[width=\columnwidth]{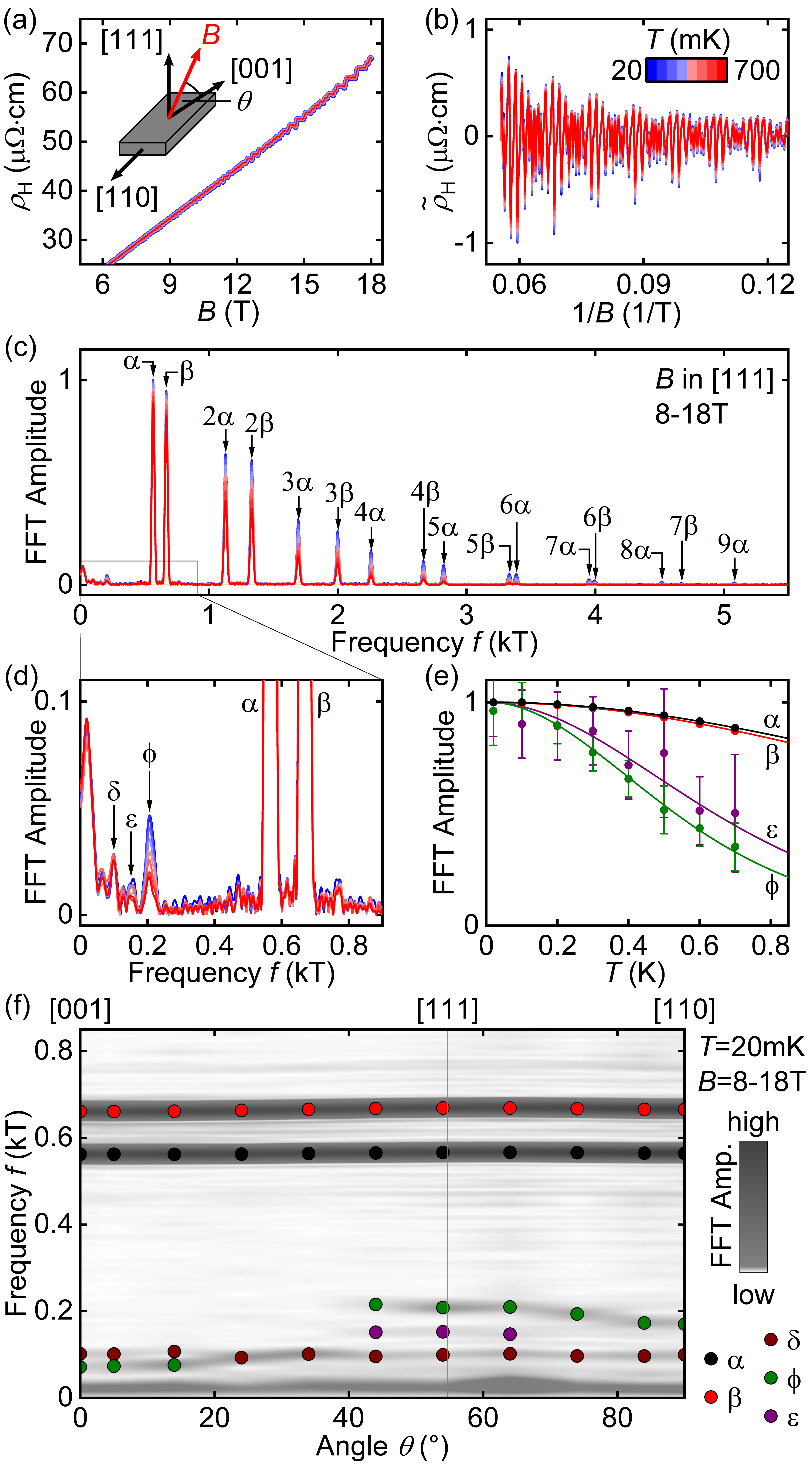}
	\caption{\label{fig:exp-SdH} Shubnikov-de Haas oscillations in CoSi. (a)~Hall resistivity $\rho_\mathrm{H}$ as a function of magnetic field $B \parallel [111]$ at different temperatures. The inset shows the experimental geometry. (b)~Oscillations in $\rho_\mathrm{H}$ as a function of $1/B$ after subtraction of a smooth background. (c)~FFT spectrum in the field range between 8 and 18\,T. The main frequencies $\alpha$ and $\beta$ and their higher harmonics are labeled. (d)~Low-frequency part of the spectrum. Three additional frequencies $\delta$, $\epsilon$ and $\phi$ are identified. (e)~Temperature dependence of the normalized FFT amplitudes. Lines show fits of the Lifshitz-Kosevich temperature reduction factor $R_T$ yielding effective masses of $m_{\alpha}=0.93\,m_{e}$, $m_{\beta}=0.99\,m_{e}$, $m_{\phi}=3.1\,m_{e}$ and $m_{\epsilon}=2.5\,m_{e}$. (f)~FFT amplitude as a function of frequency $f$ and angle $\theta$. Several frequency branches can be traced upon variation of the angle of the applied magnetic field. The extracted frequencies of each branch are marked by colored circles.
	}
\end{figure}

%%%%%%%%%%%%%%%%%%%%%%%%%%%%%%%%%%%%%%
\subsection{Shubnikov -- de Haas effect}

CoSi exhibits an almost linear Hall resistivity $\rho_\mathrm{H}$ in the field range studied. Pronounced quantum oscillations may be observed for fields exceeding $\sim6\,{\rm T}$. Shown in Figure\,\ref{fig:exp-SdH}(a) is $\rho_\mathrm{H}$ as a function of magnetic field $B\parallel$ [111]. In the temperature range between 20 and 700\,mK the nonoscillatory component of $\rho_\mathrm{H}$ remains essentially unchanged. The oscillatory part of the Hall resistivity $\tilde{\rho}_\mathrm{H}$ as a function of inverse magnetic field following subtraction of a smooth background is shown in Fig.\,\ref{fig:exp-SdH}(b). The oscillations, which are periodic in $1/B$, are dominated by a beating pattern indicating that the strongest contributions arise from two frequencies that are close to one another.

The oscillatory signal components were analyzed using a Fast Fourier Transform (FFT) in the field range between 8 and 18\,T. The resulting frequency spectrum is shown in Fig.\,\ref{fig:exp-SdH}(c). The two strong frequencies at $f_\alpha$=566\,T and $f_\beta$=668\,T were previously reported in several studies \cite{2019_Wu_ChinPhysLett, 2019_Xu_PhysRevB, 2020_Wang_PhysRevB, 2022_Guo_NatPhys, 2022_Huber_PhysRevLett, 2022_Sasmal_JPhysCondensMatter, 2023_Huber_Nature}. In our data up to nine harmonics of $f_\alpha$ and $f_\beta$ may be distinguished underscoring a high sample quality as compared to previous studies \cite{2019_Wu_ChinPhysLett, 2019_Xu_PhysRevB, 2020_Wang_PhysRevB, 2022_Sasmal_JPhysCondensMatter}. Additionally, several frequency contributions below 0.3\,kT are observed. Shown in Fig.\,\ref{fig:exp-SdH}(d) is a close-up view of the QO spectra below $\sim0.9\,{\rm kT}$, where additional frequencies are denoted by $\delta$, $\epsilon$ and $\phi$. The detection of this large number of harmonics and additional frequencies as compared to our previous studies \cite{2022_Huber_PhysRevLett, 2023_Huber_Nature} is attributed to an improved crystalline quality of the sample we used in this work, as indicated by its large residual resistivity ratio. The novel QO frequencies $f_\epsilon$ and $f_\phi$ detected are likely completely suppressed by Dingle damping in previous studies.

The frequency $f_\delta$, which corresponds to the difference between $f_\alpha$ and $f_\beta$, is due to quantum oscillations of the quasiparticle lifetime as described in Ref.\,\onlinecite{2023_Huber_Nature}. It originates in a non-linear interband coupling between the FS pockets centered at the R-point. As it is not associated with an extremal FS cross section in the spirit of the Onsager relation, it is not at the center of the work reported in this paper. We note that the amplitude of this frequency component is expected to depend on the Dingle damping factors of the constituent frequencies $f_\alpha$ and $f_\beta$ as well as the strength of the coupling between them \cite{2023_Huber_Nature, 2023_Leeb_PhysRevB} which might provide a way to learn more about the microscopic origin of the involved coupling mechanism.

In the following we focus on the new frequencies $f_\epsilon$=153\,T and $f_\phi$=208\,T. Shown in Fig.\,\ref{fig:exp-SdH}(e) are the temperature dependences of the amplitudes of $f_\epsilon$ and $f_\phi$ as well as $f_\alpha$ and $f_\beta$. The temperature dependence of $f_\epsilon$ and $f_\phi$ is described well in terms of the reduction factor $R_T$ in the LK-formalism yielding cyclotron masses of $m_{\epsilon}=2.5\,m_{e}$ and $m_{\phi}=3.1\,m_{e}$. Thus, the effective masses of $f_\epsilon$ and $f_\phi$ are significantly larger than the cyclotron masses of $f_\alpha$ and $f_\beta$, which are close to the free electron mass $\approx1\,m_{e}$.

The dependence of the FFT spectra on the angle $\theta$ between the magnetic field applied in the (110)-plane and the [001] direction is shown in Fig.\,\ref{fig:exp-SdH}(f). Frequencies at which the FFTs exhibit peaks are marked by colored circles. The peaks at $f_\alpha$ and $f_\beta$ are detected over the full angular range. They display almost no dispersion as discussed extensively before \cite{2022_Huber_PhysRevLett}. Consequently, the peak at the difference frequency $f_\delta$ is also essentially dispersionless \cite{2023_Huber_Nature}.

For $\theta<20^\circ$ additional frequencies emerge. As shown below, the upper branch corresponds to $f_\delta$ while the lower branch may be attributed to an orbit related to the frequency branch $\phi$. Between $\theta\approx20^\circ$ and $\approx40^\circ$ oscillations at $f_\alpha$, $f_\beta$ and $f_\delta$ may be discerned. For $\theta>40^\circ$ the frequency branch $\phi$ may be traced up to 90$^\circ$. In contrast, the peak at $f_\epsilon$ is only present in an angular range of about $\pm 10^{\circ}$ around the [111] direction. Additional spectral weight may be discerned in the FFT spectra at frequencies below 50\,T. This spectral weight is consistent with a low frequency reported in Refs.\,\cite{2020_Wang_PhysRevB, 2022_Sasmal_JPhysCondensMatter}, which is also present in our magnetization data (cf. Fig.\,\ref{fig:exp-dHvA-SquidVSM}). However, as the FFT spectra of our SdH data in this low-frequency region vary with the background subtraction and the field range analyzed, a specific oscillation frequency cannot be determined reliably.

\begin{figure}
	\includegraphics[width=\columnwidth]{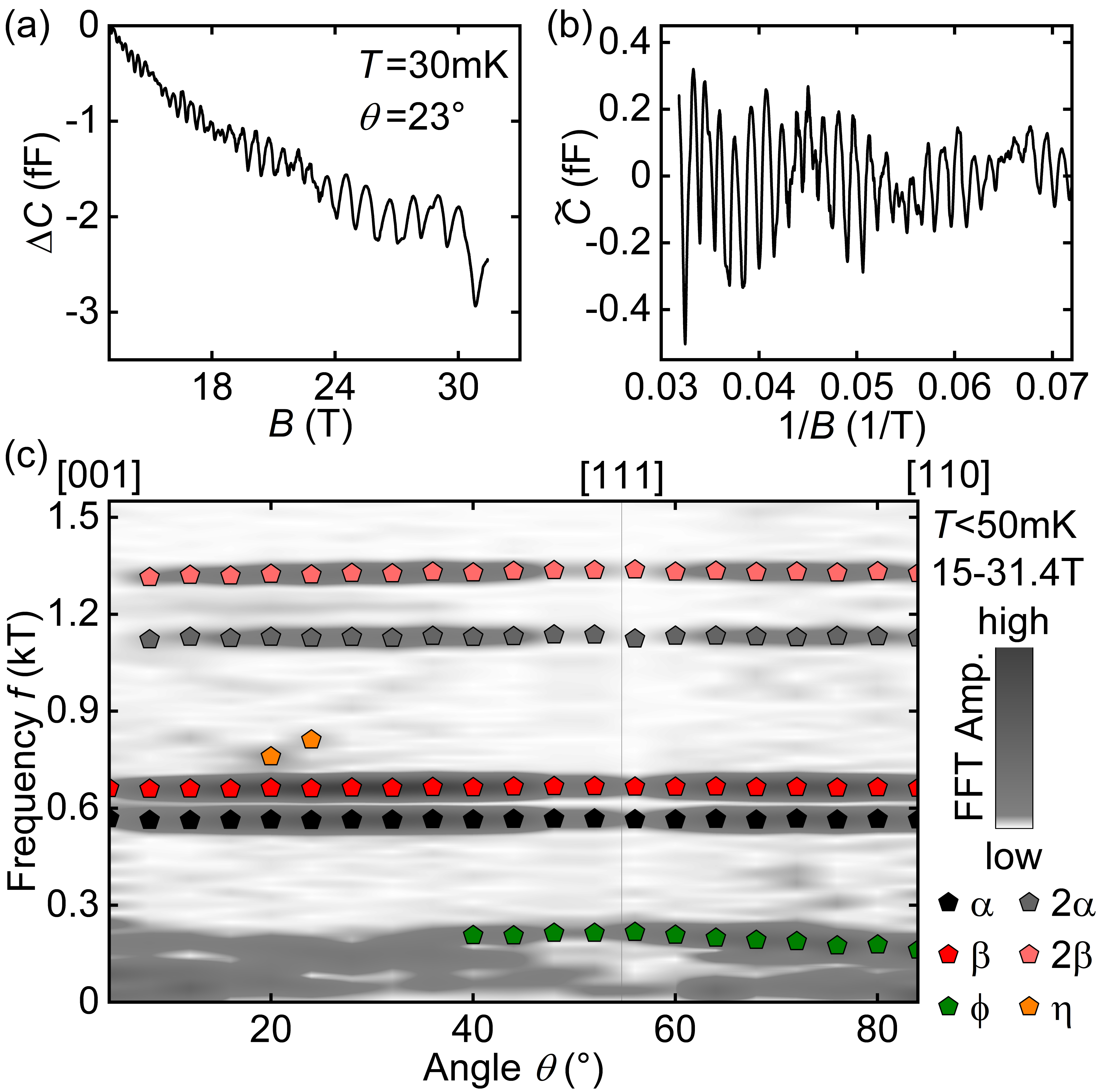}
	\caption{\label{fig:exp-dHvA-Grenoble-overview} De Haas-van Alphen oscillations in CoSi. (a)~Capacitance change $\Delta C$ of the torque magnetometer as a function of magnetic field $B$ for an exemplary data set recorded at $\theta=23^\circ$ and $T=30$\,mK. The change in capacitance is directly proportional to the magnetic torque. (b)~Oscillatory part of capacitance as a function of $1/B$ after subtraction of a smooth background. (c)~FFT amplitude as a function of frequency $f$ and angle $\theta$. In addition to the frequency branches $\alpha$, $\beta$ and their higher harmonics, the frequency branch $\phi$ can be traced for angles $>40^\circ$. In the spectra recorded at $\theta=20^\circ$ and 24$^\circ$ an additional frequency labeled $\eta$ is identified.
	}
\end{figure}

\begin{figure}
	\includegraphics[width=\columnwidth]{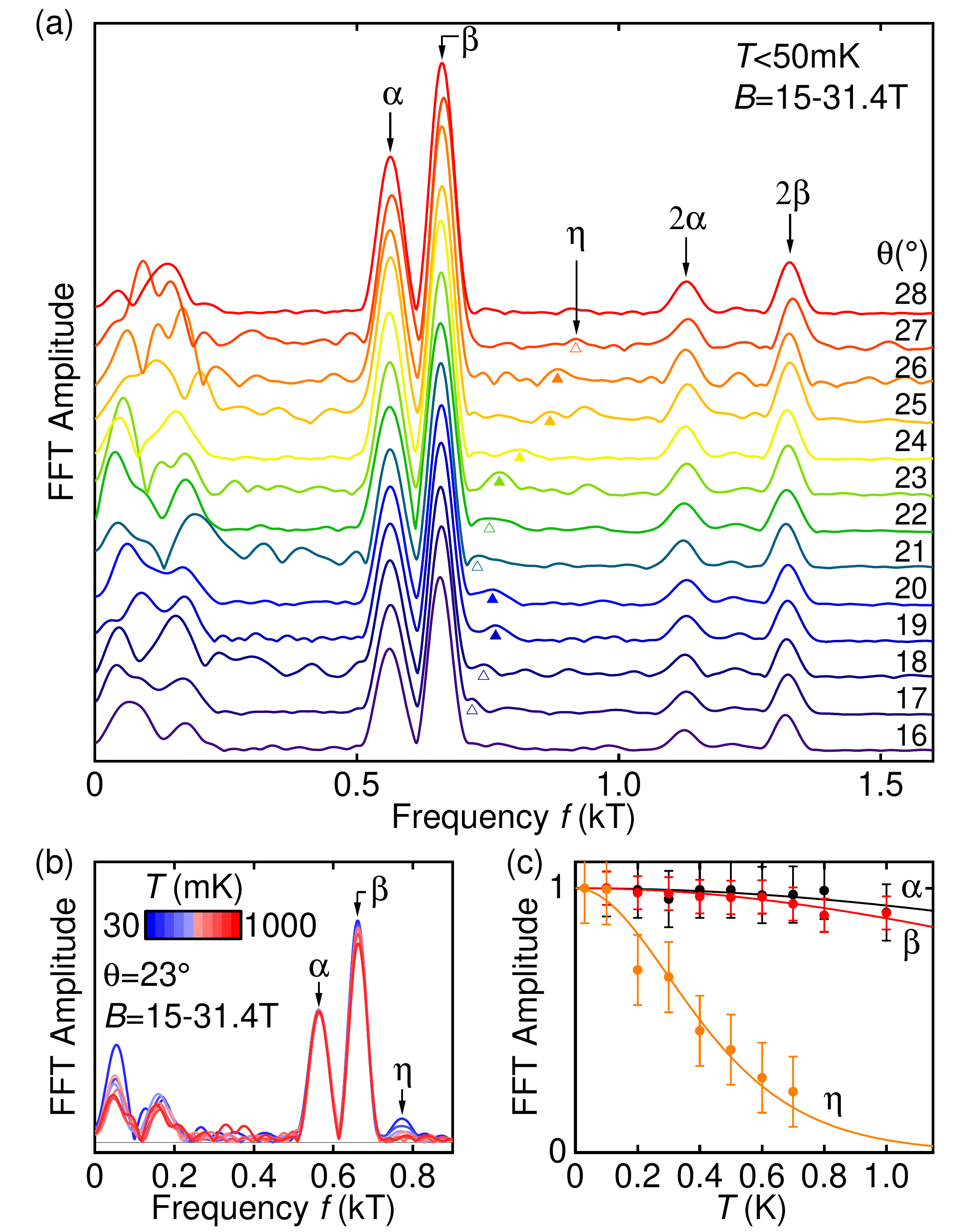}
	\caption{\label{fig:exp-dHvA-Grenoble-new-freqs} De Haas-van Alphen oscillations in CoSi in the angular range at which the frequency branch $\eta$ is identified. (a)~FFT spectra between $\theta=16^\circ$ and $28^\circ$. Curves are offset for clarity. The position of the $\eta$ peak is marked by upward triangles. Peaks satisfying all criteria defined in the Methods are indicated by solid symbols and frequencies satisfying all but one criterion by open symbols. (b)~FFT spectra at an angle of $\theta=23^\circ$ and different selected temperatures. (c)~Temperature dependence of the normalized FFT amplitudes of the identified frequencies. Lines show fits of the temperature damping factor $R_T$. The extracted cyclotron mass of $\eta$ is $m_{\eta}=7.5\,m_{e}$, while the masses extracted for $\alpha$ and $\beta$ are close to  $1\,m_{e}$.
	}
\end{figure}

%%%%%%%%%%%%%%%%%%%%%%%%%%%%%%%%%%%%%%
\subsection{de Haas -- van Alphen effect}

The de Haas-van Alphen effect was mesured using cantilever-based torque magnetometry with a capacitive readout. For a given field direction along a specific crystallographic orientation the change in capacitance is thereby proportional to the magnetic torque and thus the component of the magnetization perpendicular to the applied magnetic field. A typical data set recorded at a temperature of 30\,mK and $\theta=23^\circ$ is shown in Fig.\,\ref{fig:exp-dHvA-Grenoble-overview}(a) for magnetic fields between 15 and 31.4\,T. As for the SdH data, the oscillatory signal contributions shown in Fig.\,\ref{fig:exp-dHvA-Grenoble-overview}(b) were determined following subtraction of a nonoscillatory signal determined by a polynomial fit. The resulting frequency spectra for different angles are shown in Fig.\,\ref{fig:exp-dHvA-Grenoble-overview}(c).

Consistent with the SdH data, two almost dispersionless frequency branches denoted $\alpha$ and $\beta$ are observed, as well as their higher harmonics. Below 0.3\,kT a QO frequency at $\phi$ may be discerned between $\theta\approx 40^{\circ}$ and $\approx 90^{\circ}$. The angular dispersion of $f_\phi$ is in good agreement with the dispersion observed in the SdH spectra (cf. Fig.\,\ref{fig:comparison-exp-theory}). However, it is not possible to identify peaks at $f_\delta$ and $f_\epsilon$ in the dHvA data.

In the FFT spectra recorded at $\theta=20^\circ$ and $24^\circ$ an additional frequency with a low amplitude denoted $\eta$ may be observed between 0.7 and 0.8\,kT. To track the $\eta$ frequency in more detail the angular range between $\theta=16^\circ$ and $28^\circ$ was investigated in smaller steps as shown in Fig.\,\ref{fig:exp-dHvA-Grenoble-new-freqs}(a). The peak at $f_\eta$ could be identified at multiple angles and the strongest signal contribution was detected at $\theta=23^\circ$. Frequency contributions that satisfied all but one of the criteria defined in the methods section are marked with open symbols.

Oscillation spectra were recorded for different temperatures between 30\,mK and 1\,K at $\theta=23^\circ$ and the temperature dependence of the FFT is shown in Fig.\,\ref{fig:exp-dHvA-Grenoble-new-freqs}(b).
Figure~\ref{fig:exp-dHvA-Grenoble-new-freqs}(c) shows the FFT amplitudes as a function of temperature. While the frequencies $\alpha$ and $\beta$ decrease only slightly with increasing temperature up to $\sim$1\,K, the amplitude of $\eta$ decreases strongly  with increasing temperature and vanishes above $\sim$700\,mK. A fit of the temperature damping factor $R_T$ yields a high cyclotron mass of $m_\eta=7.5\,m_{e}$. In comparison, the cyclotron masses associated with $f_\alpha$ and $f_\beta$ inferred from the dHvA data are close to $1\,m_{e}$ in agreement with the SdH data and previous studies \cite{2019_Wu_ChinPhysLett, 2019_Xu_PhysRevB, 2022_Sasmal_JPhysCondensMatter, 2022_Guo_NatPhys, 2022_Huber_PhysRevLett, 2023_Huber_Nature}.

\begin{figure}
	\includegraphics[width=\columnwidth]{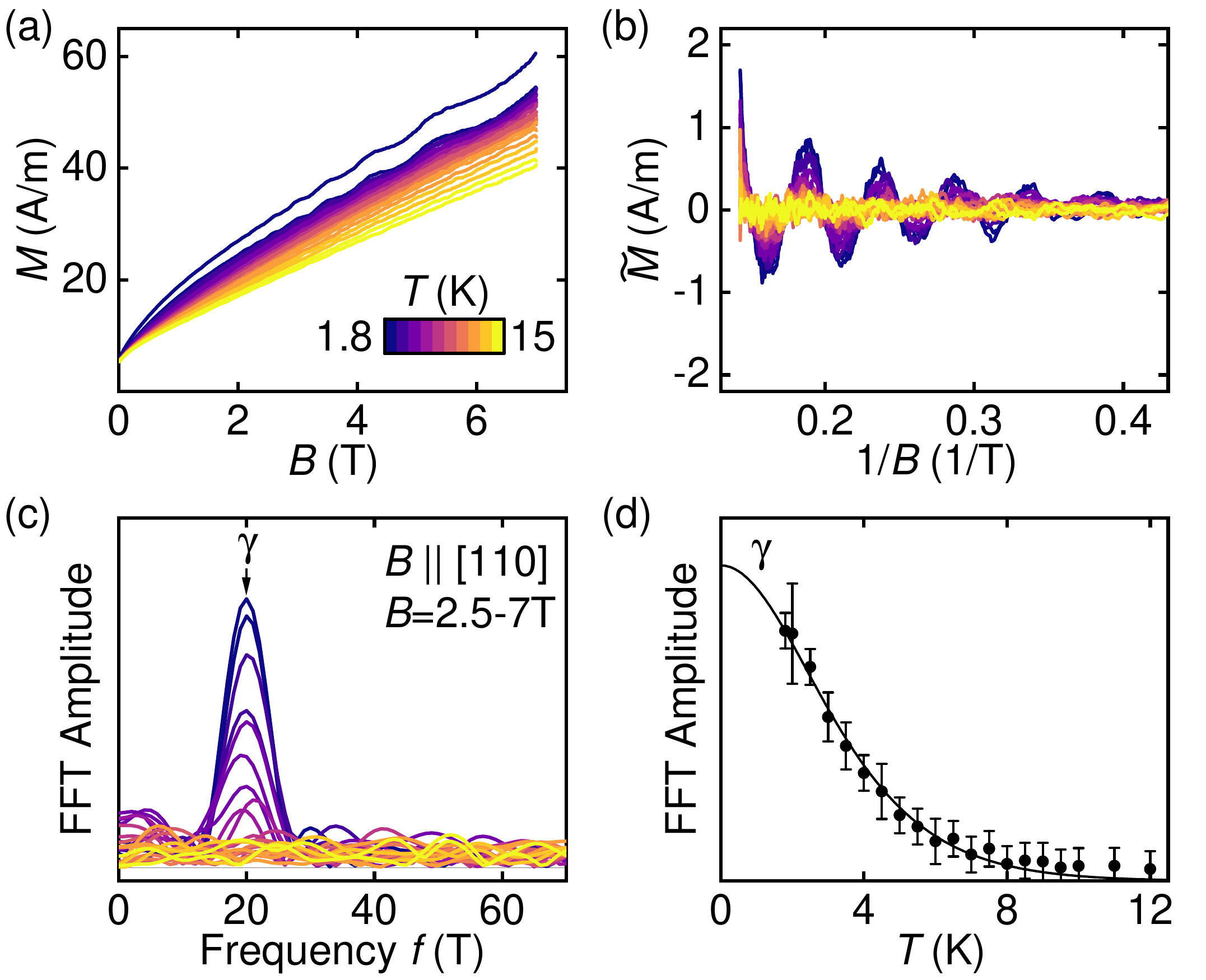}
	\caption{\label{fig:exp-dHvA-SquidVSM} De Haas-van Alphen oscillations in CoSi in the magnetic field region up to $B$=7\,T measured with a SQUID vibrating sample magnetometer~(VSM). The field is applied along the [110] direction. (a)~Magnetization $M$ versus $B$ at different fixed temperatures. (b)~Oscillatory part of the magnetization versus $1/B$ after subtraction of a smooth background. (c)~FFT spectrum of the data shown in (b) in the field range 2.5-7\,T. A low-frequency contribution at $f_\gamma=20\pm1$\,T is detected. (d)~FFT amplitude of $\gamma$ versus $T$. The line represents a fit with the LK temperature reduction factor $R_T$, yielding a cyclotron mass of $m_{\gamma}=0.17\,m_{e}$.
	}
\end{figure}

In order to resolve a low oscillation frequency reported in Refs.~\cite{2020_Wang_PhysRevB, 2022_Sasmal_JPhysCondensMatter} that could not be clearly identified in our SdH and torque magnetometry experiments, we measured the magnetization $M$ under magnetic field up to 7\,T using a SQUID magnetometer.
$M(B)\parallel [110]$ at various temperatures is shown in Fig.\,\ref{fig:exp-dHvA-SquidVSM}(a). The oscillatory signal component of the magnetization following subtraction of a monotonically increasing background is shown in Fig.\,\ref{fig:exp-dHvA-SquidVSM}(b) as a function of $1/B$, where the periodicity is clearly visible. The corresponding peak in the FFT spectrum at $f_\gamma=20$\,T is shown in Fig.\,\ref{fig:exp-dHvA-SquidVSM}(c). 

The temperature dependent decrease of the oscillation amplitude is shown in Fig.\,\ref{fig:exp-dHvA-SquidVSM}(d). It follows the LK temperature reduction factor $R_T$ with a cyclotron mass of $m_\gamma=0.17\,m_{e}$. The value observed for $f_\gamma$ is consistent with Refs.~\cite{2020_Wang_PhysRevB, 2022_Sasmal_JPhysCondensMatter} and the value for the cyclotron mass is consistent with Ref.~\cite{2022_Sasmal_JPhysCondensMatter}, but differs slightly from Ref.~\cite{2020_Wang_PhysRevB}, which reports a smaller cyclotron mass of $m_\gamma=0.11\,m_{e}$.

%%%%%%%%%%%%%%%%%%%%%%%%%%%%%%%%%%%%%%
\section{Discussion}

\begin{figure*}
	\includegraphics[width=\textwidth]{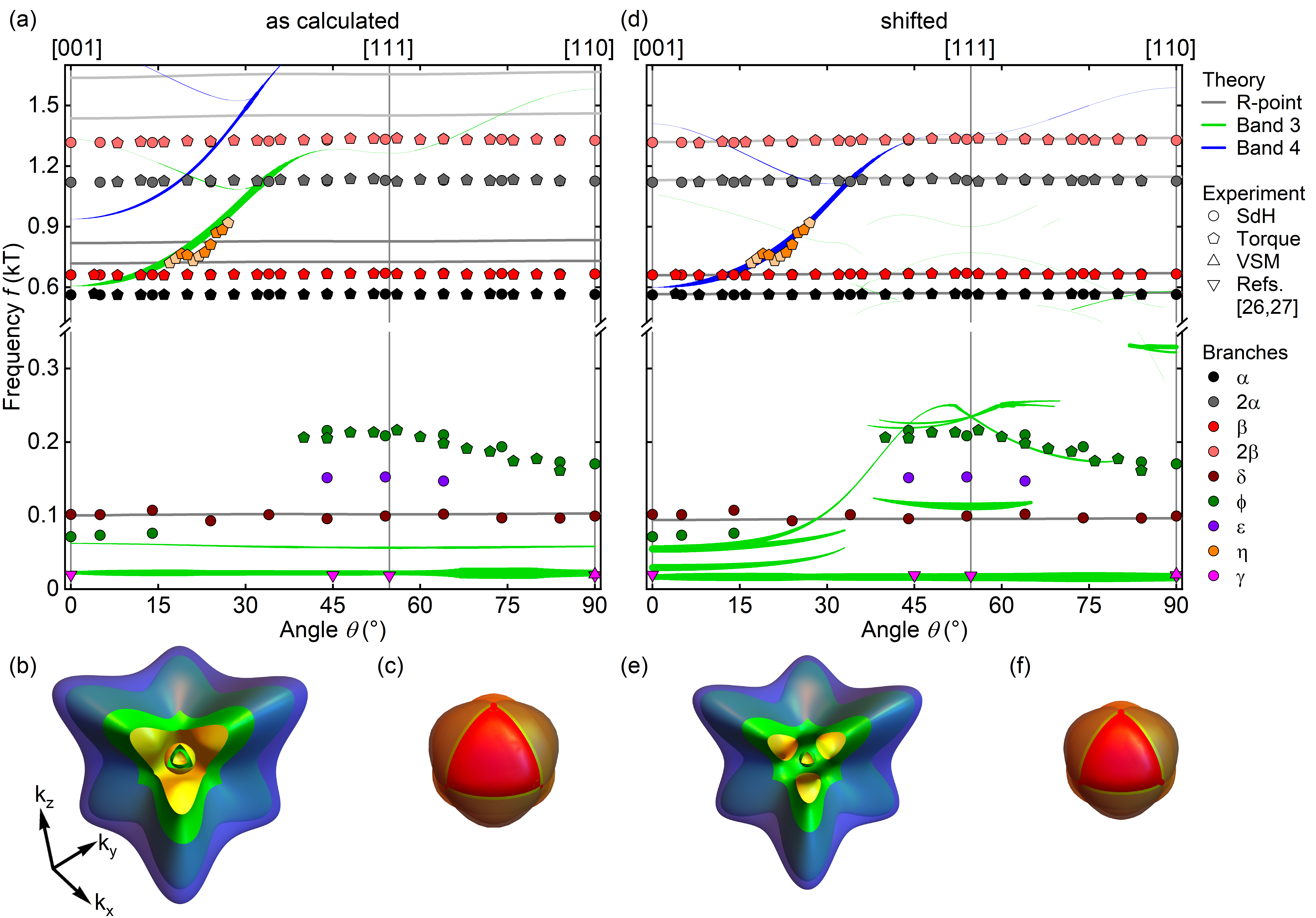}
	\caption{\label{fig:comparison-exp-theory} Comparison between calculated and experimentally detected frequency branches and FS sheets. (a) Comparison of the experimentally detected frequency branches (filled symbols) with the DFT predictions (colored lines). Symbol shapes denote the technique used. Different colors represent different frequency branches. The light orange symbols of the $\eta$-branch correspond to the open symbols in Fig.~\ref{fig:exp-dHvA-Grenoble-new-freqs}. Angle-dependent VSM data on the low-frequency branch $\gamma$ are included from Refs.~\cite{2020_Wang_PhysRevB, 2022_Sasmal_JPhysCondensMatter}. The line color indicates which band gives rise to the extremal orbit and the linewidth represents the expected amplitude. (b)  Close-up view of the FS pockets around the $\Gamma$-point as calculated by DFT. (c) FS pockets around the R-point as calculated. (d) Frequency branches as in (a), but after applying small rigid band shifts to the DFT results. A close match to experiment is achieved for all branches detected experimentally. (e),(f) FS pockets around $\Gamma$ and R, respectively, as matched to experiment.
	}
\end{figure*}

In the following, we compare the experimental results with the DFT calculations in order to determine the FS geometry of CoSi. We begin with a comparison between experiment and DFT as calculated, followed  step by step by the considerations that permit a unique and unambiguous assignment. On this note, it is helpful to recall that DFT represents, in principle, a parameter-free technique. However, DFT relies on approximate functionals that do not fully capture the effects of correlations. As a result, the calculated and the real band dispersions may differ in a way that is not fully understood, asking for an empirical correction. As shown below, in CoSi the calculated FS and the experimental results differ moderately. Small rigid shifts of the bands may therefore be used to match the experimental and calculated quantum oscillation frequencies when determining the FS.

Shown in Fig.\,\ref{fig:comparison-exp-theory}(a) are the angular dispersions of all QO frequencies detected experimentally together with the oscillation frequencies predicted by the DFT as calculated. The linewidths of the calculated branches are scaled with the expected QO amplitudes in order to highlight the relative weight of different branches as described above. For what follows, it is helpful to note that the majority of those predicted frequency branches that were not observed experimentally are expected to contribute only weakly to the overall signal.

In addition to the dispersion of a given branch, its cyclotron mass represents another important indicator for matching the experimentally detected QOs with predictions from first-principles calculations. The experimentally determined and calculated cyclotron masses for selected field orientations are summarized in Table~\ref{tab:effective-masses}. 

\begin{table}[]
\centering
\caption{\label{tab:effective-masses} Experimentally determined and calculated cyclotron masses at specific angles $\theta$.}
\begin{tabular}{|c|c|c|c|}
	\hline
		Frequency branch & $m_\mathrm{exp}$ ($m_e$) & $m_\mathrm{calc}$ ($m_e$) & Field angle $\theta$ ($^\circ$) \\ \hhline{|=|=|=|=|}
		$\gamma$         & 0.17                     & 0.13                      & 90 [110]                        \\ \hline
		$\epsilon$       & 2.5                      & 2.6                       & 54.7 [111]                      \\ \hline
		$\phi$           & 3.1                      & 4.3                       & 54.7 [111]                      \\ \hline
		$\eta$           & 7.5                      & 7.2                       & 23                              \\ \hline
	\end{tabular}
\end{table}

%%%%%%%%%%%%%%%%%%%%%%%%%%%%%%%%%%%%%%
\subsection{R-centered FS pockets}

We start the assignment of the QO frequencies to the FS sheets with the two almost dispersionless frequency branches $\alpha$ and $\beta$, as denoted in Fig.\,\ref{fig:comparison-exp-theory}(a) together with their second harmonics. Both frequencies were previously observed in several studies \cite{2019_Wu_ChinPhysLett, 2019_Xu_PhysRevB, 2020_Wang_PhysRevB, 2022_Guo_NatPhys, 2022_Huber_PhysRevLett, 2022_Sasmal_JPhysCondensMatter}, where they were independently and unambiguously assigned to the four Fermi surface pockets around the R-point. Their flat angular dispersion and the pairwise degeneracy in cross-sectional area is explained by a combination of spin-orbit coupling, symmetry-enforced nodal plane degeneracies on the BZ boundaries and complete magnetic breakdown at near-degeneracies \cite{2022_Huber_PhysRevLett, 2022_Guo_NatPhys}. The corresponding calculated branches shown as gray lines match the flatness of the dispersion perfectly, but are offset towards higher frequencies by roughly 160\,T. We conclude that the calculated electron pockets around R depicted in Fig.\,\ref{fig:comparison-exp-theory}(c) are larger than in reality. Empirically shifting the bands at the R-point up by 30\,meV yields a perfect match between the experimental and theoretical branches a shown in Fig.\,\ref{fig:comparison-exp-theory}(d). This establishes that the real FS at the R-point is approximated well by the four sheets depicted in Fig.\,\ref{fig:comparison-exp-theory}(f). With this, the R-point pockets are fully determined, since no further branches are expected.

The size of the R-pockets determined experimentally has important consequences for the possible band shifts around $\Gamma$. The required shift at R reduces the Fermi volume occupied by the electron pockets from $2$\,\% as calculated to about $1.5$\,\% in the real material. As CoSi is semi-metallic with Fermi surface sheets at R and $\Gamma$ only, charge neutrality implies that the electron pockets must be compensated by  predominantly hole-like pockets at the $\Gamma$-point. In turn, a band-shift downward is expected at $\Gamma$ on the order of 7\,meV, taking into account the higher density of states due to the larger band masses as compared to the bands at R. 

The above argument based on charge neutrality assumes the absence of unintentional doping. This is justified by considering the literature \cite{2019_Wu_ChinPhysLett, 2019_Xu_PhysRevB, 2020_Wang_PhysRevB, 2022_Guo_NatPhys, 2022_Huber_PhysRevLett, 2022_Sasmal_JPhysCondensMatter} as well as the work presented here, where values of $f_\alpha$ and $f_\beta$ differ by less than $\pm 10$\,T across many samples grown by different methods in different laboratories, corresponding to a variation of less than $\pm 0.1$\,\% of the Fermi volume.

%%%%%%%%%%%%%%%%%%%%%%%%%%%%%%%%%%%%%%
\subsection{Difference frequency}

As mentioned above, the dispersionless $\delta$ branch at the semi-classically forbidden difference frequency $\beta-\alpha$ observed in the transverse magnetoresistance and the Hall effect can be explained by QOs of the quasiparticle lifetime due to non-linear interband coupling between the FS pockets around the R-point \cite{2023_Huber_Nature, 2023_Leeb_PhysRevB}. If there is a coupling between two distinct FS orbits, the lifetime of quasiparticles on these orbits oscillates with the frequencies associated with both orbits. This may give rise to combination frequencies in the QO spectra. Since they are not related to an extremal cross section via the Onsager relation, the combination frequencies do not reveal any new information about the FS geometry. However, they might provide a tool to gain information about the nature of interactions between quasiparticles on different parts of the Fermi surface and thereby give insight into a material's properties beyond its FS geometry. Because the goal of this work is the determination of the FS, we focus on the conventional quantum oscillation frequencies in the following.

%%%%%%%%%%%%%%%%%%%%%%%%%%%%%%%%%%%%%%
\subsection{$\Gamma$-centered FS sheets}

The branches $\phi$, $\eta$, $\epsilon$ and $\gamma$ observed experimentally arise from the  FS sheets centered at $\Gamma$. Taking into account small rigid band shifts, all four frequencies may be assigned unambiguously based on matching frequency, angular dispersion, angular range in which they occur, cyclotron mass, angle dependent oscillation amplitude, and the consistency with charge neutrality within the accuracy discussed above. 
We start out by first describing the frequency branches corresponding to DFT as calculated.

Shown in Fig.\,\ref{fig:comparison-exp-theory}(b) are the FS sheets at $\Gamma$, which arise from bands 3 and 4. Band 3 as calculated leads to multiple sheets of which the innermost is a spherical hole pocket that would give rise to a dispersionless frequency branch at 22\,T [lowest green branch in Fig.\,\ref{fig:comparison-exp-theory}(a)]. It is surrounded by an almost spherical electron pocket corresponding to a frequency branch at $\approx$60\,T [middle green branch in Fig.\,\ref{fig:comparison-exp-theory}(a)]. 

The outermost hole sheet arising from band 3 is cuboid-shaped with elongations along the $\Gamma$-R-directions exhibiting one minimal and two degenerate maximal extremal cross sections for $B\parallel [001]$ (Note that the intermediate electron pocket can be understood as a "hole" in the middle of the outermost hole sheet). QOs arising from the branch of the minimal cross section at about $0.6$\,kT are predicted to be much stronger than the ones arising from the maximal branch at $1.3$\,kT, due their respective masses and curvature factors. Upon rotation towards [111] the minimal and maximal orbits merge and form a single low-amplitude frequency branch [upper green branches in Fig.\,\ref{fig:comparison-exp-theory}(a)]. 

The calculated FS arising from band 4 consists of a single pocket whose shape resembles the outermost sheet of band 3. Accordingly, the resulting frequency branches shown in blue in Fig.\,\ref{fig:comparison-exp-theory}(a) resemble those of the outermost sheet of band 3 displaced by $\approx$300-400\,T towards higher frequencies.

No further branches are expected for the FS geometry as calculated, which is in clear contradiction to experiment. Namely, the experimental branches in the $70-220$\,T regime, $\phi$ and $\epsilon$, cannot be explained by the FS as calculated in DFT. Conversely, DFT predicts a dispersionless branch at $\approx$60\,T which is absent in experiment.

In comparison, the experimentally observed branches can be described consistently in DFT when shifting band 3 down by $4$\,meV and shifting band 4 down by $7$\,meV, resulting in the FS geometry shown in Fig.\,\ref{fig:comparison-exp-theory}(e). We note that both the sign and size of the rigid band shift applied at $\Gamma$ are consistent with the band shift applied at the R point.

Due to this small shift, the intermediate electron pocket of band 3 merges with the outer hole pocket forming a multiply connected combined sheet that features apertures along the $\Gamma$-X directions. Nested inside this is the small spherical hole pocket arising from the same band. This change in FS geometry explains the branches $\epsilon$ and $\phi$.

\begin{figure}
	\includegraphics[width=\columnwidth]{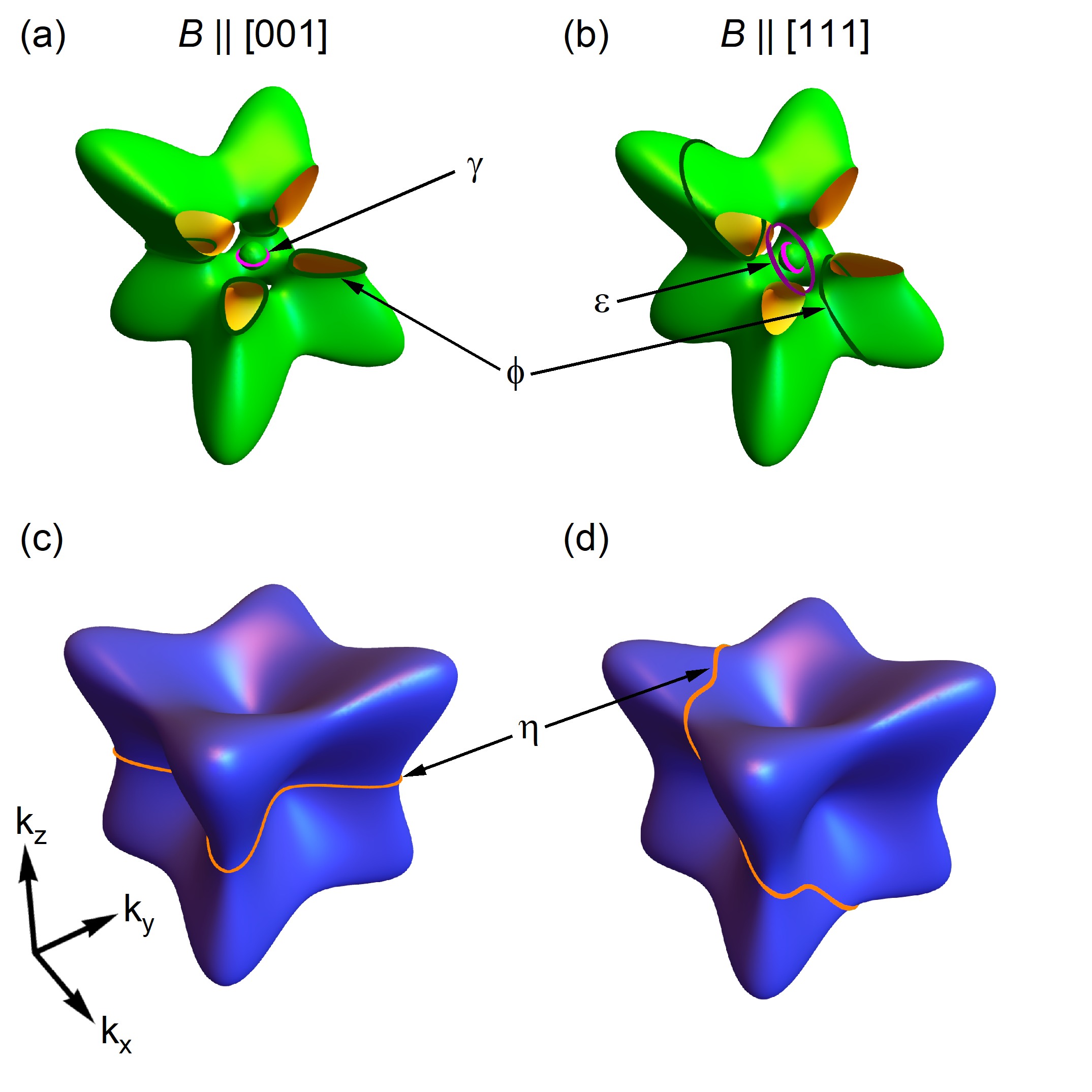}
	\caption{\label{fig:FS-extremal-orbits} Extremal orbits on the FS sheets of CoSi around the $\Gamma$-point. (a, b) FS arising from band 3 in Fig.~\ref{fig:band structure-and-fs} shifted downwards by 4\,meV. Part of the outer pocket has been cut out to allow for a better view. The $\gamma$ branch is assigned to the small hole-like sphere in the center. The $\phi$ orbits thread through the apertures of the outer pocket, whereas the $\epsilon$ orbit runs on the inner surface of the same pocket and exists in a limited angular range only.  (c, d) Fermi surface originating from band 4 in Fig.~\ref{fig:band structure-and-fs} shifted downwards by 7\,meV. The $\eta$ branch can be linked to the extremal orbit with minimal cross section.}
\end{figure}

The $\phi$ branch with its distinct angular dispersion can be identified as arising from the new orbits around the "necks" created by the shift. These orbits pass through the apertures in the outer FS sheet of band 3, as shown in Fig.\,\ref{fig:FS-extremal-orbits}(a,b) for two field directions.
This branch is detected in both the SdH and dHvA data over the full angular range, with the exception of the strongly dispersive part between $\theta=20^{\circ}$ and $\theta=40^{\circ}$, where the calculated orbit looses most of its spectral weight. We note that there are possible contributions in this frequency range in our dHvA data shown in Fig.~\ref{fig:exp-dHvA-Grenoble-overview}(c) which are, however, not conclusive.
For $\theta<20^\circ$, the experimental branch resides at $\approx$70\,T, and reemerges at angles larger than $\theta$=40$^\circ$ where it exhibits a moderate dispersion with frequencies varying between $f_\phi=221$\,T for $B\parallel [111]$ and $f_\phi=170$\,T for $B\parallel [110]$. 
This behavior is matched well by the DFT calculations after the downward shift by $4$\,meV and can be understood as the neck-orbits for $B\parallel [001]$ in Fig.\,\ref{fig:FS-extremal-orbits}(a) passing over the "lobes" of the pocket when rotating the field towards [110], with a maximal cross section close to the [111] direction.

In addition, the cyclotron mass was analyzed for field applied along the [111] direction, being about 25\,\% smaller than estimated from DFT [see Table~\ref{tab:effective-masses}]. However, we note that close to the [111] direction the calculated value of $m_\phi$ is subject to a large uncertainty both with respect to the angle and with respect to the exact band shape. In turn, this represents a satisfactory match.

The $\epsilon$-branch is observed experimentally in a narrow angular range around the [111] direction only featuring a weak angular dispersion.
As shown in Fig.~\ref{fig:FS-extremal-orbits}(b), it originates from an orbit running along the "inside" of the outer FS sheet of band 3. For other orientations, e.g., for field applied along  [001] as shown in Fig.~\ref{fig:FS-extremal-orbits}(a), this orbit does not exist because of the apertures along the $\Gamma$-X-lines. For the angular range in which this branch is expected to exist, the dispersion and the cyclotron mass  [see Table~\ref{tab:effective-masses}] match the experimentally detected $\epsilon$-branch well. 

The dispersionless low-frequency branch denoted $\gamma$ was reported before in dHvA experiments recording directly the magnetization \cite{2020_Wang_PhysRevB, 2022_Sasmal_JPhysCondensMatter}, where it was attributed to a small pocket at $\Gamma$. In Ref.~\onlinecite{2022_Sasmal_JPhysCondensMatter} $f_\gamma$ was not assigned to a specific band, whereas Ref.\,\onlinecite{2020_Wang_PhysRevB} assigned $f_\gamma$ to band 1. However, a band shift of about $45$\,meV upward would be required for this to be correct, resulting in a dramatic hole doping in excess of $10$\,\% of the Fermi volume. In turn, an assignment to band 1 can clearly be ruled out.

The extremal orbit of the innermost spherical pocket of band 3 matches the $\gamma$ branch for a downward shift by 4\,meV, reducing the frequency from $22$\,T as calculated, to $\approx$20\,T as observed experimentally. The almost perfect isotropy of this pocket naturally explains why it evades detection in the torque magnetometry we used. 
While in principle also bands 1, 2 and 6 could give rise to light and fairly spherical pockets, only the assignment to band 3 is consistent with the assignment of the other branches. In particular an assignment to band 1 or 2 would imply pocket dimensions of band 3 and 4 that are much too large to fit the observation of  branch $\eta$. In addition, such an assignment would raise the question why other light pockets required by the band symmetry for such a shift are not observed. Along the same line, an assignment to band 6 can be ruled out, because this would require that no outer sheet of band 4 is present that could host the $\eta$ orbit.

For the downward shift of band 3 by 4\,meV the intermediate electron pocket surrounding the small inner hole pocket merges with the large outer hole pocket. This naturally explains the absence of a branch from the intermediate electron pocket, which should otherwise be straightforward to detect due to its small cross sectional area and small mass. 

$f_\eta$ is, finally, observed in our high-field dHvA experiments over a limited angular range around $\theta=23^\circ$.
The frequency branch $\eta$ is consistent with the minimal orbit on the FS pocket arising from band 4 [Fig.~\ref{fig:FS-extremal-orbits}(c and d)]. Its frequency matches theory when shifting band 4 down by 7\,meV. The shift of band 4 being slightly larger than the shift required for band 3 may be explained in two ways. Either band 4 follows the dispersion of band 3 a bit more closely than determined in DFT, or the "waist" of this pocket is more pronounced than determined in DFT. 
The presence of $\eta$ in such a narrow angular range reflects, most likely, a combination of different factors. First, with a calculated cyclotron mass ranging between $6$ and $12$\,$m_e$, it is by far the heaviest orbit observed reducing the amplitude of the QOs in terms of the cyclotron energy and the temperature reduction factor $R_T$. Second, the $\eta$-branch is predicted to have large weight only in the region where it is observed experimentally. Apart from the cyclotron mass which increases with increasing angle up to a maximum for the field applied along the [111] direction, this may be attributed to the torque factor $\frac{1}{f}\frac{df}{d\theta}$, present in our high-field dHvA experiments and a stronger Dingle damping as the orbit becomes larger with increasing angle.

%%%%%%%%%%%%%%%%%%%%%%%%%%%%%%%%%%%%%%
\subsection{Conclusions}

We conclude that the Fermi surface of CoSi closely resembles the depiction shown in Fig.\,\ref{fig:comparison-exp-theory}(e) and (f). In particular the sheets of band 3 around $\Gamma$ do not consist of three nested pockets (or rather, an outer pocket with a hole inside, in which a small inner pocket is situated), but one inner spherical pocket and one outer multiply connected sheet with six apertures along the $\Gamma$-X directions. Both the R-centered electron pockets and the $\Gamma$-centered predominantly hole-like sheets are smaller than predicted by DFT, consistent with global charge neutrality. Differences between experiment and theory may be ascribed to the limitations of DFT and be avoided by means of more accurate calculations as performed, e.g., in Ref.\,\cite{2018_Pshenay-Severin_JPhysCondensMatter}. Knowing the FS geometry and dimensions, we can now comment on the relative positions of the topological crossings and their relevance for topological response functions.

\begin{figure}
	\includegraphics[width=\columnwidth]{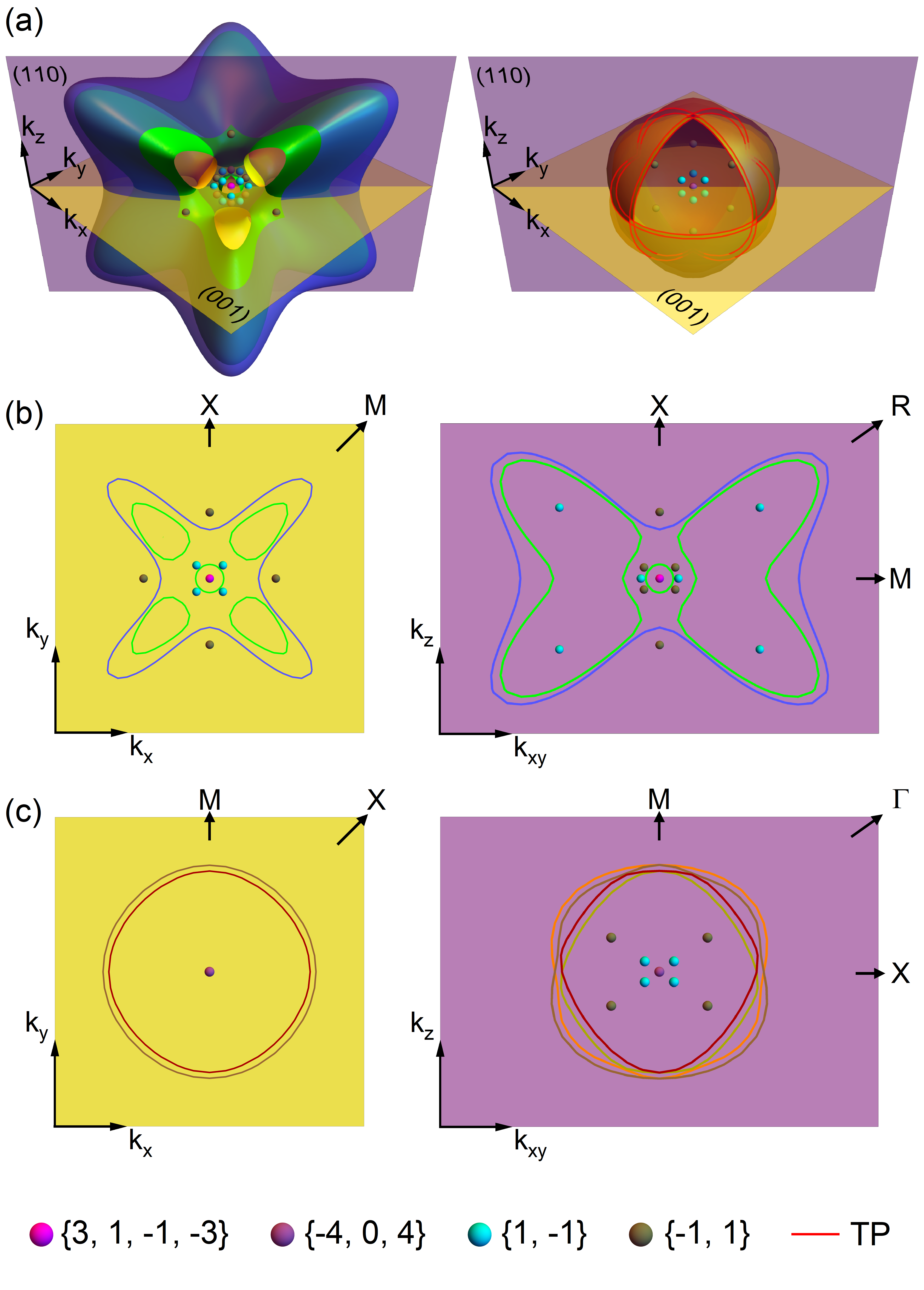}
	\caption{\label{fig:FS-vs-crossing-points} Positions of topological band crossings with respect to the FS. Crossing points are shown as colored dots where the color represents the charge of the involved bands sorted by increasing band index. Red lines indicate topological protectorates (TPs) on nodal planes (cf. Refs.~\cite{2021_Wilde_Nature, 2022_Huber_PhysRevLett}) (a)~Perspective views of the FS sheets centered on $\Gamma$ and R. (b)~FS contours and band crossings on high-symmetry planes around the $\Gamma$-point as indicated in (a). (c)~FS contours and band crossings in high-symmetry planes around the R-point.}
\end{figure}

Shown in Fig.~\ref{fig:FS-vs-crossing-points} is the distribution of topological crossings in the vicinity of the $\Gamma$ and the R pockets. Around the $\Gamma$ point, shown in Fig.~\ref{fig:FS-vs-crossing-points}(a), the innermost spherical pocket of band 3 encloses only the 4-fold crossing, such that the net signed Berry curvature flow through the surface is +3 (note that the signs depend on the handedness of the enantiomorph of CoSi). The Weyl points along $\Gamma$-M and $\Gamma$-R close to this innermost pocket are not surrounded by a closed Fermi surface. Their locations may be seen more clearly in Fig.~\ref{fig:FS-vs-crossing-points}(b), where the point crossing locations (dots) and the FS contours in the $\Gamma$-centered (001) and (110) planes are shown. Note that these Weyl points would be enclosed by a FS in case of unshifted bands, where they would add up to a net topological charge of $(-1\times12+1\times8)=-4$. The multiply connected outer sheet of band 3 encloses Weyl points on the $\Gamma$-R-line with a topological charge $(1\times8)=8$. The outermost enclosing sheet arising from band 4 encloses the multifold point at $\Gamma$ and the Weyl points on $\Gamma$-R contributing a net charge of $(1\times1-1\times8)=-7$.

For the sake of completeness, we briefly summarize the well-established result for the four R-centered pockets arising from bands 5,6,7 and 8. Here, the situation is complicated by the trios of topological nodal planes covering the entire Brillouin zone boundary. These nodal planes carry total topological charges of $\{-5,5\}$ for bands 5,6, and $\{-1,1\}$ for bands 7,8. They form topological protectorates on the Fermi surface sheets [red lines in Fig.~\ref{fig:FS-vs-crossing-points}(a)], as the crossings are pinned to the FS. However, the charge contributed by the nodal planes enclosed in the FS pockets is not quantized. In addition, the charges of the multifold crossing at R cannot be evaluated using the Abelian Berry curvature, since no integration contour can be chosen that is gapped everywhere, due to the presence of the nodal planes. Instead, the non-Abelian Berry curvature~\cite{2018_Pshenay-Severin_JPhysCondensMatter} allows one to calculate the Chern numbers representing the topological charges of band pairs $(3,4)$, $(5,6)$ and $(7,8)$, yielding $\{-4,0,4\}$ \cite{2022_Huber_PhysRevLett}. Thus, we obtain $(1\times0-1\times8)=-8$ for charge enclosed by the FS pair $(5,6)$, because of two Weyl points on $\Gamma$-R (one crossing between band 5 and 6 compensating itself and one crossing with band 7) and $(1\times8+1\times4)=12$ for the FS pair $(7,8)$ due to the Weyl point on $\Gamma$-R and the multifold point at R.

The main purpose of our study concerned the detailed and complete experimental determination of the FS of CoSi. In principle, based on our results, the relation of the FS to the topological crossings may now be determined unambiguously. However, an analysis of the topological charges as performed above, is not only interesting from a fundamental point of view but may also provide important insights when searching for topological contributions in the physical properties.

Response functions, such as CPGE~\cite{2021_Ni_NatCommun}, the intrinsic spin Hall effect~\cite{2016_Sun_PhysRevLett}, and the quantum non-linear Hall effect~\cite{2015_Sodemann_PhysRevLett}, in general anti-correlate with the distance of the crossings to the Fermi energy and Fermi surface, since the Berry curvature entering in these response functions is maximal at the crossings. Due to this connection, establishing the relation between band crossings and the Fermi surface is an important step forward.

We wish to point out that the net flow of Berry curvature through FS pockets does - as a rule - not fully determine the topological responses per se apart from idealized special cases (e.g., the quantized CPGE \cite{2017_deJuan_NatCommun,2020_Rees_SciAdv}). Rather, the integration "kernels" for different response functions incorporate the Berry curvature in different non-trivial ways. In general, this makes it necessary to perform the relevant integrations explicitly. The resulting Berry curvature will depend sensitively on the exact positions of the bands with respect to the Fermi energy as determined in our study.
For example, Ref.\,\cite{2021_Ni_NatCommun} calculates the CPGE assuming a band shift of $17$\,meV in the opposite direction as compared to the band shift determined experimentally in our study, corresponding to a strong overall hole doping resulting in FS sheets around $\Gamma$ that differ strongly from the real ones. Using the band shifts reported here, such calculations may now be refined. 

Further, the quantization of the CPGE is very sensitive to the chemical potential, and the multifold points at $\Gamma$ and R contribute most to this response. Since the position of the multifold point at $\Gamma$ derived here is closer to the Fermi energy, we expect a broader photon energy range of CPGE quantization stemming from the $\Gamma$ point than previously expected. For the same reason, we expect topological responses originating from the multifold crossing at $\Gamma$ to be larger in general.

Explicit numerical calculations of response functions based on the experimentally refined Fermi surface will naturally take into account Berry curvature contributions arising from the multiple hotspots of Berry curvature discussed here as well as from near "quasidegeneracies," which do not lead to a net flux through a closed surface, but may nevertheless contribute to response functions \cite{2004_Yao_PhysRevLett,2022_Guo_NatPhys,2022_Wilde_NatPhys}. When trying to describe this network of Berry curvature contributions with \emph{a posteriori}  constructed effective Hamiltonians, a host of such Hamiltonians valid in different k-space regions would be necessary, limiting the value of this approach for the calculation of response functions.

In conclusion, we report SdH and dHvA measurements in CoSi investigating the angular dispersion of quantum oscillations and analyzing the temperature dependence of the oscillation amplitudes at specific angles. We identify a total of six frequency branches, three of which exhibit large cyclotron masses. These frequency branches allow the determination of the FS shape of CoSi and its relation to the topological band crossings. This way our findings may facilitate a quantitative understanding of topological response functions in CoSi.

\begin{acknowledgments}
	We gratefully acknowledge the support of the LNCMI-CNRS, member of the European Magnetic Field Laboratory (EMFL) as well as financial support of DFG via TRR360 - 492547816 (ConQuMat), SPP 2137 (Skyrmionics) under grant no. PF393/19 (project id 403191981), DFG-GACR project WI3320/3-1 (project id 323760292), ERC Advanced grant no 788031 (ExQuiSid)ExQuiSid), and DFG cluster EXC 2111 Project No. 390814868.
\end{acknowledgments}

%%%%Bibliography%%%%

%


\begin{thebibliography}{50}
\bibitem [{Com()}]{CommentMFP}%
	\BibitemOpen
	\href@noop {} {\bibinfo  {journal} {We note that this mean free path is only
			a rough estimate from the angular dependence of the oscillation amplitudes.
			Fits of the Dingle damping factor to the magnetic field evolution of the
			amplitudes were not feasible for the $\Gamma$-centered
			orbits}\ }\BibitemShut {NoStop}%
\end{thebibliography}
\end{document}